# A Functional Human Liver Tissue Model: 3D Bioprinted Co-culture Discoids

**Vignesh Subramaniam[1,†], Carolina Abrahan[2,†], Brett Higgins[3], Steven J. Chisolm[1], Baleigh Sweeney[1], Senthilkumar Duraivel[4], Leandro Balzano-Nogueira[5], Glyn D. Palmer[2,*], Thomas E. Angelini[1,6,7,*]**

1 Department of Mechanical and Aerospace Engineering, Herbert Wertheim College of Engineering, University of Florida, Gainesville, Florida, United States of America

2 Department of Orthopaedic Surgery and Sports Medicine, College of Medicine, University of Florida, Gainesville, Florida, United States of America

3 Department of Molecular Medicine, Morsani College of Medicine, University of South Florida, Tampa, Florida 33612, United States of America

4 Department of Materials Science and Engineering, Cornell University, Ithaca, New York, United States of America

5 Department of Pathology, Immunology and Laboratory Medicine, Diabetes Institute, College of Medicine, University of Florida, Gainesville, Florida, United States of America.

6 Department of Materials Science and Engineering, Herbert Wertheim College of Engineering, University of Florida, Gainesville, Florida, United States of America

7 J. Crayton Pruitt Family Department of Biomedical Engineering, Herbert Wertheim College of Engineering, University of Florida, Gainesville, Florida, United States of America

† The authors contributed equally to this work.

* Corresponding authors. Email: t.e.angelini@ufl.edu, palmegd@ortho.ufl.edu

## Abstract

To reduce costs and delays related to developing new and effective drugs, there is a critical need for improved human liver tissue models. Here we describe an approach for 3D bioprinting functional human liver tissue models, in which we fabricate disc-shaped structures (discoids) 200 μm in thickness and 1-3 mm in diameter, embedded in a highly permeable support medium made from packed microgels. We demonstrate that the method is precise, accurate, and scalable; up to 100 tissues per hour can be manufactured with a variability and error in diameter of about 4%. Histologic and immunohistochemical evaluation of printed discs reveal self-organization, cell cohesion, and key liver marker expression. During the course of 3-4 weeks in culture, the tissues stably synthesize albumin and urea at high levels, outperforming spheroid tissue models. We find the tissues express more than 100 genes associated with molecular absorption, distribution, metabolism, and excretion (ADME) at levels within the range of human liver. The liver tissue models exhibit enzymatic formation of metabolites after exposure to multiple test compounds. Together, these results demonstrate the promise of 3D printed discoids for pharmacological and toxicological applications.

## 1. Introduction

Neither *in vitro* tissue models nor animal models are currently able to predict and prevent drug induced liver injury (DILI), leading to marketed pharmaceutical withdrawals and clinical development failures[1], yet they are widely used in pharmaceutical and biopharmaceutical discovery and development[2-6]. Among these models, two-dimensional (2D) monolayers of iPSC-derived human hepatocytes and bioengineered micro-patterned co-cultures of primary hepatocytes and fibroblasts have exhibited some morphological and functional features of *in vivo* liver tissue[7, 8]. However, the limitations of monolayer cultures in predicting liver-specific responses are widely recognized, despite their frequent use throughout the pharmaceutical industry[9-13]. For example, primary hepatocytes cultured in 2D monolayers tend to de-differentiate and rapidly lose liver-specific functions[14-16]. To overcome the deficits of 2D approaches, numerous 3D culture systems have been developed; it is believed that the formation of cell–cell, and cell–matrix interactions in 3D are better at maintaining cell activity and function than in 2D cultures[2, 17]. Three-dimensional (3D) approaches include spheroid culture of primary hepatocytes[18], 3D bioprinted liver tissue in trans-wells[19], and 3D printed scaffolds seeded with hepatocytes[10, 20]. These 3D systems are usually benchmarked by functional characteristics of human liver *in vivo*, such as albumin and urea production and ADME gene expression profiles[21].

One way to improve the functional performance of tissue models is to provide perfusion; liver tissue models cultured in microfluidic perfusion chambers exhibit higher viability, oxygen saturation, and synthesis rates of albumin and urea than their static counterparts[22]. To simulate physiological fluid and solute transport, culture media has been driven through 3D populations of hepatocytes sandwiched between collagen gels[23]. Toward the goal of making large-scale engineered tissue constructs, advanced methods for creating vasculature have been developed[24, 25]. While developing a diversity of advanced engineering approaches has helped the field move closer to the goal of producing functional tissue models, the complexity of both the manufacturing methods and the fluidic systems prevent their broad adoption in practice and create scale-up challenges in industrial applications. These challenges are not encountered with simpler approaches to manufacturing tissue models, like spheroid culture, but such simple approaches produce tissue models that lack critical functions and exhibit high variability, preventing accurate Critical Quality Attributes (CQA) quantification of the final products[1]. Overall, there remains a serious need in the pharmaceutical industry for a functional liver tissue model having minimal barriers to adoption; the functional tissue must be manufactured reproducibly and at relatively high throughput and maintained without complex perfusion equipment.

Here, we describe a method for rapidly and precisely manufacturing functional human liver tissue models that can be cultured and assayed for more than 21 days without complex fluidic systems or specialized apparatus. To create millimeter-scale constructs that do not need to be perfused, we manufacture disc-shaped structures (discoids) that are 200 μm thick. The tissues are 3D bioprinted directly into a medium made by packing soft granular-scale spheres (microgels) together, swollen in cell growth media. The biofabricated method is precise and accurate, with less than 5% variability and error in the effective diameter of printed discoid tissues. The biofabricated liver tissue discoids exhibit stable albumin and urea synthesis at target levels, ADME gene expression profiles that mimic the human liver, and the ability to metabolize test compounds. We fabricate the discoids directly into the wells of 96-well plates at a rate exceeding one tissue per minute. This approach is compatible with standard biolaboratory equipment and instrumentation, while also able to be scaled-up and integrated with product quality assessments in future commercial manufacturing efforts.

## 2. Results

To create a versatile support medium that exhibits minimal interactions with cells, liquid media, and any other 3D bioprinted materials like extracellular matrix (ECM), we leverage a polymer that is commonly used to prevent cell and protein adhesion to surfaces: polyethylene glycol (PEG)[26]. We synthesize spherical PEG hydrogel particles by emulsifying the precursor mixture and performing a crosslinking reaction while stirring (see Methods). After the reaction is complete, we perform several cycles of washing followed by sterilization and a fluid exchange with cell growth media. 100 μL of the microgel media is deposited into each well of a 96-well plate and incubated before 3D printing. We prepare various mixtures of hepatocytes, endothelial cells, and collagen-1 and load them into sterile syringes that are mounted onto our printer (described later). The printer is programmed to follow a spiral path while extruding the cell mixture, printing 96 tissues in approximately 1 h. When printing is complete, 100 μL of liquid media is gently pipetted into each well on top of the packed microgels, which remain settled at the bottom of the wells (Fig. 1). The tissues are incubated and assayed for up to 21 days, as described in the following sections.

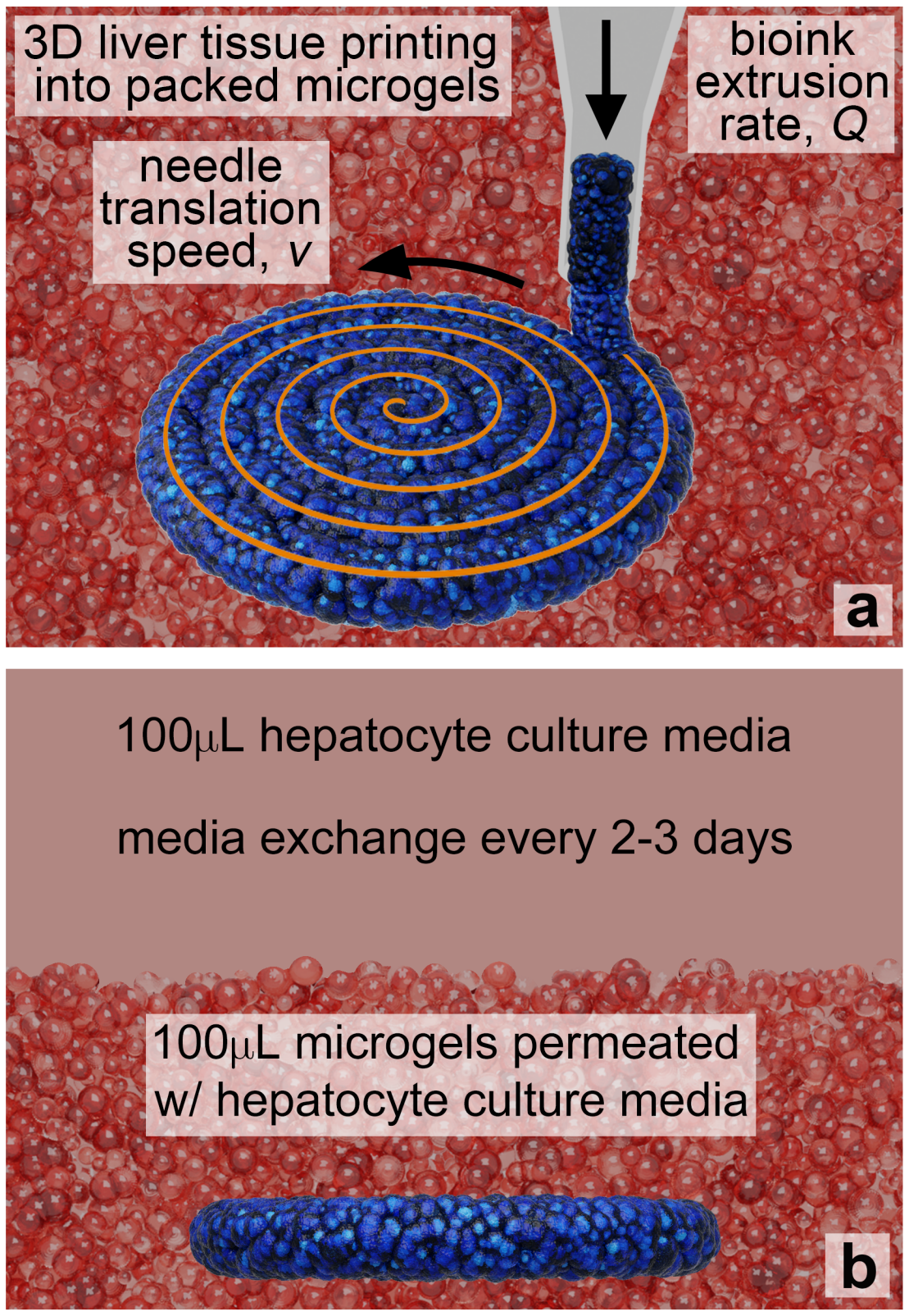

**Figure 1: 3D bioprinting of liver tissues into PEG microgel media.** (a) An illustration of 3D printing liver tissues in PEG microgels. The bioink extrusion rate, $Q$, and the needle translation speed, $v$, were controlled to fabricate thin liver tissue discoids. (b) A two-dimensional illustration of the 3D printed liver tissue in microgel media with a fluid supernatant in a single well of a multi-well culture plate. 100μL of hepatocyte culture media was exchanged every 2-3 days according to the hepatocyte manufacturer protocol.

## 2.1 PEG microgel media formulation for 3D bioprinting

To formulate packed microgel media for 3D printing high-quality liver tissue models, the medium must easily flow under moderate levels of applied stress, it must exhibit an elastic modulus high enough to support printed tissues, and it must be made from particles of the right size range. For example, if the average microgel particle radius is sub-micrometer in scale, the particles will be thermally active, and the medium will exhibit fluid-like properties and flow over long timescales. However, if particles are too large, their physical size will limit the quality of printing. To quantify the size distribution of PEG microgels synthesized as described in Methods section 4.1, we diluted the gels in PBS buffer and collected images with an inverted brightfield microscope (Fig. 2a,b). To measure the diameters of hundreds of microgels, a standard edge detection algorithm is employed to segment and isolate all individual objects in each image. Non-circular objects are discarded, and the projected area of all remaining objects are measured. Equating each measured area to that of a circle and solving for the corresponding radius, $R$, we construct a probability density function of particle radius, $p(R)$. We find that $p(R)$ follows a log-normal distribution. The median diameter of the microgel particles was found to be 6.26 ± 0.18 µm (± 95% confidence interval)[27]. This size range is consistent with previous work on high-quality 3D bioprinting into microgel media[28, 29].

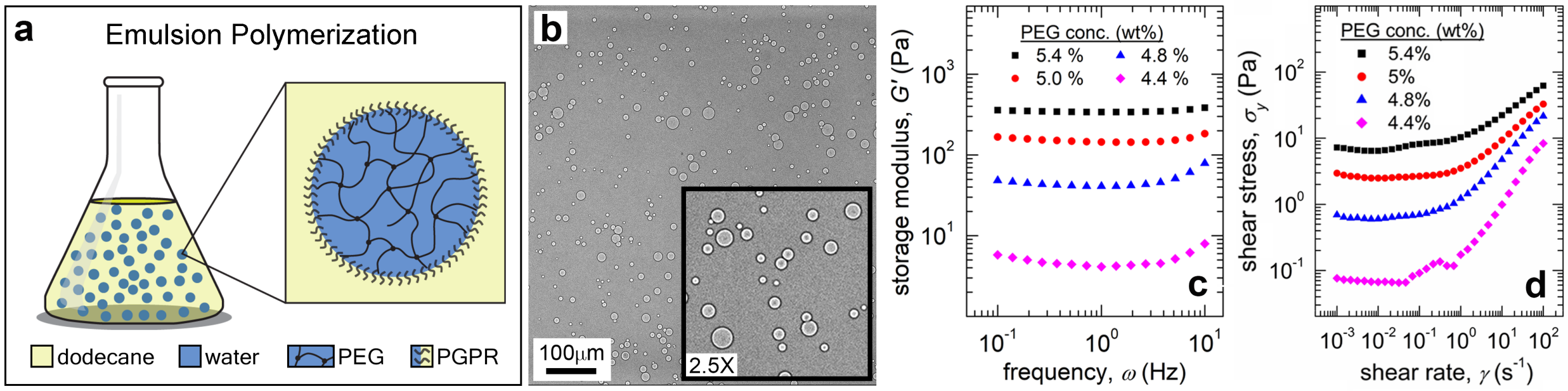

**Figure 2: PEG Microgel Synthesis and Characterization.** (a) Polyethylene glycol (PEG) microgels were synthesized using inverse emulsion polymerization. (b) PEG microgels exhibited circular cross-sections with a mean diameter (± standard deviation) of 9.46 ± 4.25 µm, as shown here with a representative brightfield microscopy image. (c) Elastic shear moduli of microgel solutions (G') are nearly frequency independent over the tested frequency range of 0.1-10 Hz and greater than viscous moduli (G") for a wide range of microgel concentrations, indicating that the material behaves like a damped elastic solid under low levels of shear. (d) The material exhibits low yield stress in the range of 0.1-10 Pa at concentrations above loose packing, and the yield stress of the material is linearly proportional to the elastic shear modulus.

To choose a formulation that exhibits the flow and elasticity properties needed for 3D bioprinting, rheological tests were performed on microgel dispersions prepared at a variety of concentrations using an Anton Paar MCR 702 rheometer with 25 mm and 50 mm parallel plate geometries. We performed small amplitude frequency sweeps to determine the elastic and viscous shear moduli in the linear deformation regime. A strain amplitude of 1% was applied to the samples oscillating over a frequency range of 0.1 – 10 Hz. The microgels exhibit nearly frequency-independent moduli over the entire tested frequency range across concentrations (Fig. 2c). Elastic shear moduli are larger than viscous moduli at PEG concentrations above 4.2% (*w*/*w*), demarcating a crossover concentration where microgels transition from liquid-like state to a damped elastic solid-like state at low levels of shear; below this concentration the measured torque values fall below the instrument's sensitivity [27]. To investigate the yielding of microgel packs under persistent shear, we performed unidirectional shear tests, ramping the shear rate from 0.001 to 100 $s^{-1}$ while measuring the shear stress. We find the shear stress curves exhibit plateaus at low shear rates corresponding to the different materials' yield stresses, falling within the approximate range of 0.1-10 Pa (Fig. 2d). Guided by these rheological measurements, we chose to formulate the microgel medium at a PEG concentration of 5% (*w/w*) for 3D liver tissue models, which yield at a stress of 2 Pa during the printing process but provide an elastic modulus of *G'* = 100 Pa, sufficient to support the deposited structures as we found with other microgel systems[30-33].

**2.2 Bioprinting into PEG microgels produces structures of predictable size and shape**

We performed a series of tests designed to explore the level of control over the size and shape of thin disc-shaped 3D printed liver tissue models (discoids). To create sufficiently large structures that can be assayed and cultured in the porous PEG microgel medium without the need for perfusion, we designed 200 µm thick discoid tissue models with diameters between approximately 1 mm and 3 mm. The discoid structures were created using planar spiral print paths. In these tests, we used fluorescently labeled HepaRG hepatocytes mixed with cell growth media and collagen-1 as the 3D bioprinting ink (Methods section 4.2). The bioprinted structures were imaged using an inverted Nikon Eclipse Ti-2 laser scanning confocal microscope with a C2+ scan head to evaluate their quality (Fig. 3a-d). Maximum intensity projections were taken along the optical axis of each z-stack. To determine the projected area, *A*, of the printed tissues, the maximum intensity projections were thresholded and segmented, and the areas of the largest detected objects were computed. To measure the effective spiral diameters, $d_m$, we performed a contour integral along the theoretical spiraling backbone of a feature having a width that matched the spiral pitch, *c*. The resulting area is given by

$$A = \frac{\pi}{4}\left( {d_m}^2 - c d_m + \frac{c^2}{4} \right),$$

where $d_m$ is the distance from the center of the spiral to the endpoint on the edge. This formula enables the determination of $d_m$ from a measurement of *A*, given by the relationship

$$d_m = \frac{c}{2} + 2\sqrt{\frac{A}{\pi}}.$$

Thus, the effective diameter of the spiral is equal to the diameter of an equivalent circle plus one-half the spiral pitch. To compare $d_m$ to expected diameter, $d_e$, we account for the width of the feature in a simple formula for a spiral, given by

$$d_e = fc\left(\frac{\phi}{\pi} + \frac{1}{2}\right),$$

where $\phi$ is the total accumulated polar angle corresponding to a spiral of diameter $d_e$, and $f = 1.1$ is an empirically determined calibration factor. Plotting $d_m$ versus $d_e$ for different spiral diameters and replicates, we found an average error in diameter of 4.2% with a standard deviation in the errors of 0.82%. The largest error in diameter across all printed tissues was 10.9% (Fig. 3e).

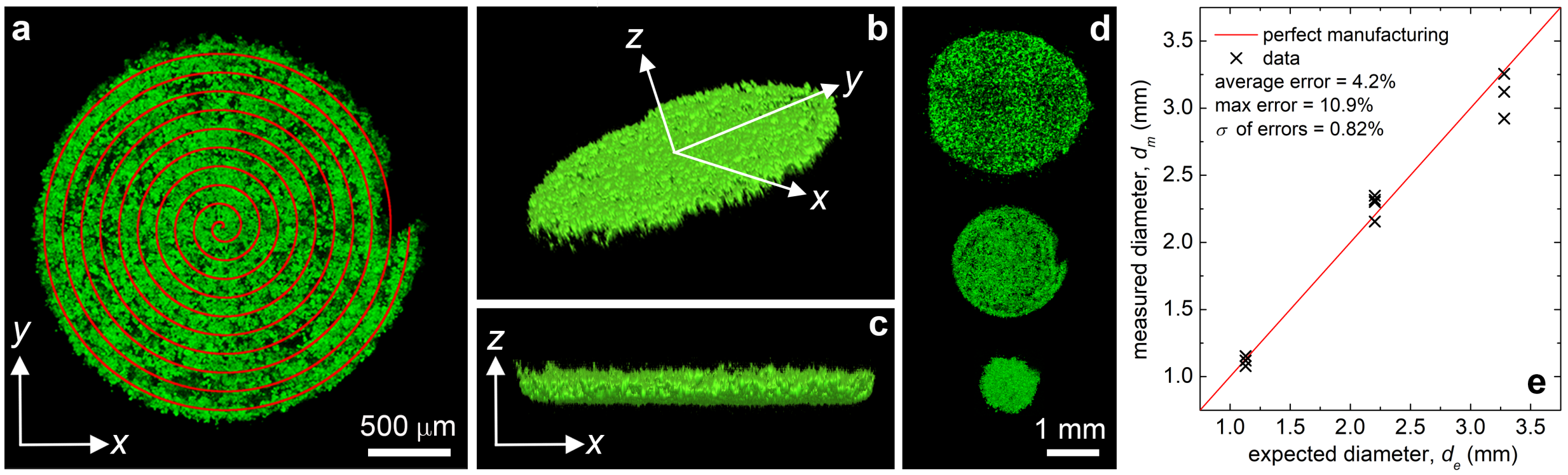

**Figure 3: Bioprinting Quality Characterization.** We printed fluorescently dyed hepatocytes into PEG microgel media to test our printed structure quality. (a) Confocal imaging of the structures revealed well-defined thin disc-shapes (discoids) resulting from the prints. The print path is illustrated by the red spiral. (b, c) The discoid shape and tissue thickness can be seen by examining the printed structures from various angles. (d) Discoids of a range of diameters were printed to test control over the structure size, shape, and precision. (e) To assess discoid manufacturing quality, the measured tissue diameter, $d_m$, is compared to the expected diameter, $d_e$. We found an average error in diameter of 4.2% with a standard deviation in the errors of $\sigma = 0.82\%$ (averaged over 12 tissues). The largest error in diameter across all printed tissues was 10.9%.

## 2.3 HUVEC co-culture enhances primary hepatocyte functions in 3D printed discoids

Albumin and urea were selected as biomarkers for baseline assessment of hepatocellular functionality and the overall health status of primary hepatocytes cultured in PEG microgels. Human liver tissue model discoids were 3D printed, composed of either primary hepatocyte monocultures (H), hepatocytes and HUVECs (H:Hu), hepatocytes and primary cholangiocytes (H:C), or hepatocytes combined with both HUVECs and cholangiocytes (H:C:Hu). The average cell number within each group did not significantly change over the culture duration, although some differences between groups were observed at D3 and D7 (range: ≈ 20,000 – ≈ 45,000 per construct) (Fig. 4a). Importantly, albumin and urea synthesis rates varied with cellular composition, where H:Hu exhibited significantly higher levels compared to H, H:C and H:C:Hu at D7 and D10 ($P < 0.05$; Fig. 4b,c), indicating that HUVECs have a stimulatory effect on albumin and urea synthesis by hepatocytes. Conversely, the incorporation of cholangiocytes, with or without HUVECs, was inhibitory to synthesis levels relative to hepatocyte monocultures (H), lowering albumin production after D3 ($P < 0.05$, Fig. 4b) and urea after D7 ($P < 0.05$; Fig. 4c).

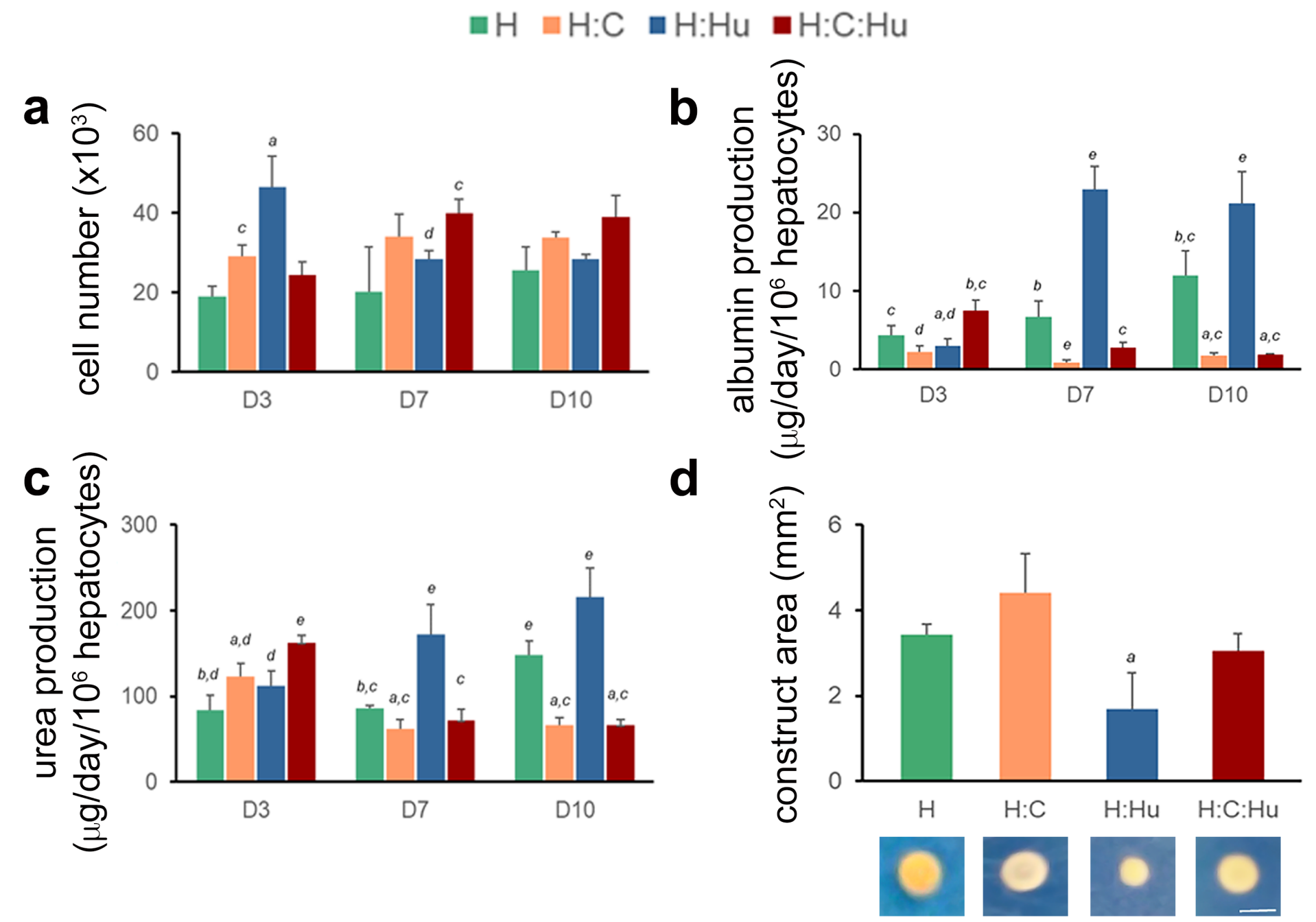

**Figure 4: Effects of HUVEC and cholangiocyte co-culture on the functional activity of 3D printed liver discoids.** (a) Mean cell number per printed construct following culture in PEG microgel media. (b) Albumin synthesis rates, adjusted to hepatocyte cell number for each culture group. (c) Urea synthesis rates, adjusted to hepatocyte cell number for each culture group. Values are mean + SD per timepoint ($n$ = 4-5). (d) Mean area of printed constructs ($n$ = 4-5 per group) following 10 d culture in PEG microgel media. Images below show representative discoids for each culture group. Scale bar = 1 mm. *H*: Primary hepatocyte monoculture; *H:C*: Hepatocyte – Cholangiocyte co-culture (2:1); *H:Hu*: Hepatocyte-HUVEC co-culture (2:1); *H:C:Hu*: Hepatocyte-Cholangiocyte-HUVEC co-culture (2:0.5:0.5). *a-e;* indicate statistical significance ($P <$ 0.05) between cell culture groups for each timepoint. *a*: significance vs H; *b*; significance vs HC; *c*; significance vs H:Hu; *d*: significance vs H:C:Hu; *e*: significance vs all other groups.

HUVEC incorporation also led to a significant ≈1.5-fold reduction in discoid size relative to the other cell groups ($P < 0.05$ vs H; Fig 4d), with the tissue models appearing more rigid compact than other cell groups after 10 days of microgel culture. These findings suggest that HUVEC incorporation provides a support structure for primary hepatocytes, which may enhance cell cohesion and hepatocellular functions.

### 2.4 Tissue model size and geometry affect hepatocellular function

Due to their finite modulus and low yield stress, PEG microgels form a porous, granular matrix that supports living cells in a defined space. Combined with the precision of our 3D printing method, this support medium permits the assembly and subsequent culture and characterization of cellular structures of controllable size and geometry. To investigate the impact of construct dimensions on hepatocellular functions, discoids of increasing diameter (1.5 – 2.5 mm) were generated from the same Hepatocyte/HUVEC/collagen preparation using separate print codes and cultured in microgel media for 3 weeks. To generate discoids of varying size, the spiral print trajectory paths were altered and programmed as described in section 2.2 As expected, cell number per construct varied proportionally with print size, and this was maintained over a 21-day culture (Fig. 5a). Print size did not significantly change albumin synthesis rates (Fig. 5b). However, urea production was higher in the 1.5 mm group with significance at 7 days ($P < 0.0005$ vs 2 mm; $P < 0.0005$ vs 2.5 mm), 14 days ($P < 0.0005$ vs 2 mm; $P < 0.0005$ vs 2.5 mm) and 21 days ($P < 0.05$ vs 2.5 mm) (Fig. 5c).

To assess the potential impact of geometric shape on microtissue viability and hepatocyte function, microtissue constructs were 3D printed as discoids and spheroids; spheroid cultures formed by self-aggregation of cells under low attachment conditions are commonly used in 3D models. Discoids and spheres (both 1.5 mm diameter) were generated from a single hepatocyte/HUVEC/collagen preparation and cultured in PEG microgels for 14 days. Consistent with their larger volume, sphere cultures contained about 4-fold greater cell numbers than discs, but were less stable over time, with a 46% decrease at 14 days (vs 7 days) compared to 33% for disc cultures (Fig. 5d). Albumin and urea synthesis rates were also considerably lower in sphere cultures relative to disc for both timepoints ($P < 0.0005$; Fig. 5e,f), indicating reduced hepatocyte function.

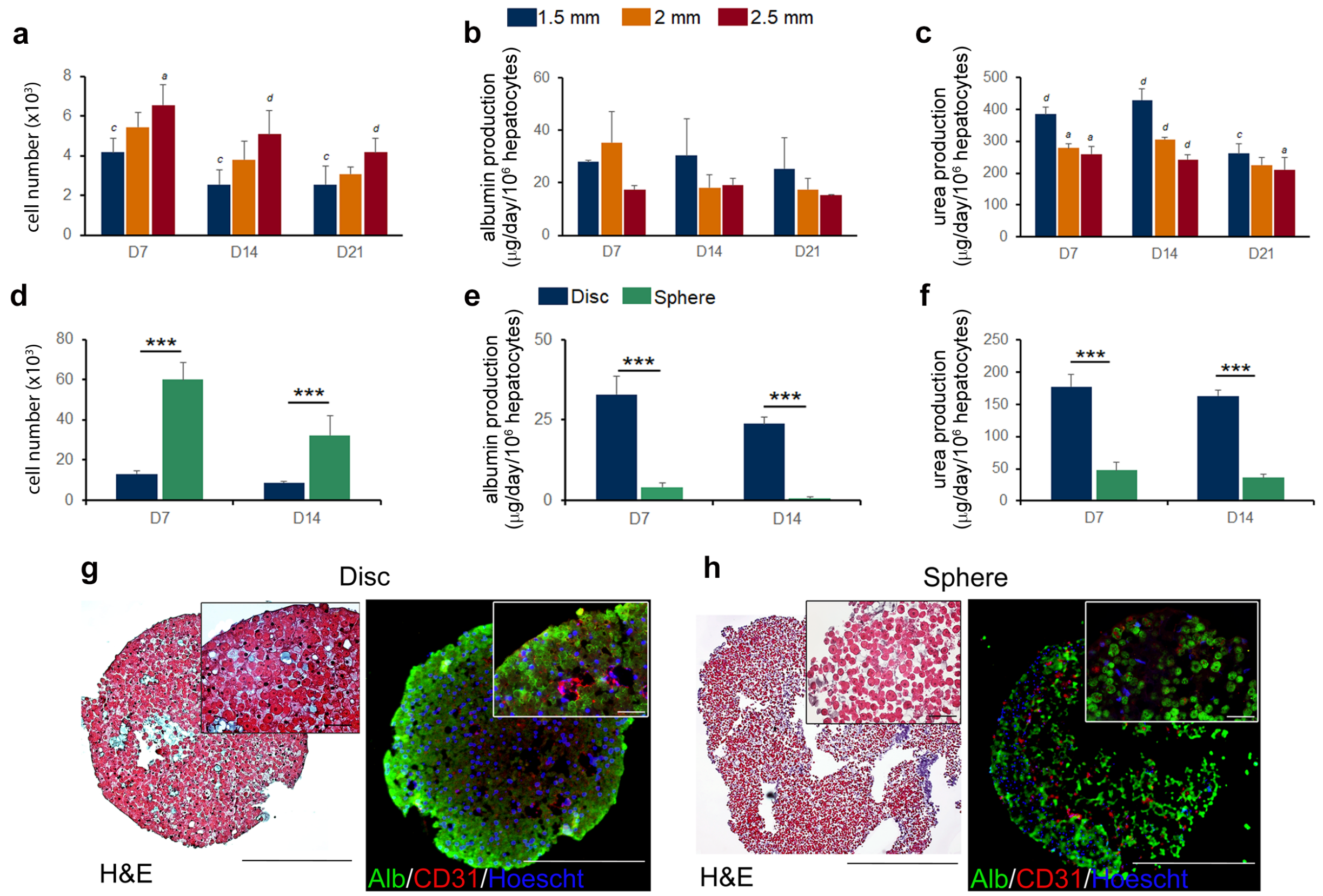

**Figure 5: Functional activity of 3D printed liver microtissues of varying size and geometry.** (a) Mean cell number per construct, (b) albumin and (c) urea synthesis rates in Hepatocyte:HUVEC co-cultures printed as discs of varying diameter (1.5 - 2.5 mm). Values are mean + SD for each timepoint/group ($n$ = 3-4) a-d; indicate statistical significance (P < 0.05) between cell culture groups for each timepoint. a: significance vs 1.5 mm group; b; significance vs 2 mm group; c; significance vs 2.5 mm group; d: significance vs all other groups. (d) Mean cell number per construct, (e) albumin and (f) urea synthesis rates in Hepatocyte:HUVEC co-cultures printed in disc or sphere geometries. Values are mean + SD for each timepoint/group ($n$ = 4-5). *** $P < 0.001$. (g) Disc and (h) Sphere H&E and Alb/CD31 immunofluorescent staining after 14 days in LLS culture. Scale bar: 500 µm (*inset panel*: 50 µm)

Histological evaluation of microtissues harvested at 14 days also revealed marked differences in overall structure and cell organization between the disc and sphere print geometries (Fig. 5g,h). Within discs, hepatocytes (rounded cells) were clustered together, with high levels of cell-cell contact. Immunofluorescence staining revealed albumin positive hepatocytes and CD31 positive HUVECs, which could be found along the construct perimeter and internally, forming luminal structures (Fig 5g). In sphere constructs, albumin, and CD31 positive cells were also evident, but the constructs were less compact with little cell-cell cohesion (Fig 5h). Together, these findings indicate geometry has a profound impact on hepatocyte function with 3D printed microtissues enabling greater contraction and cell aggregation within the tissue constructs, which is in turn accompanied by enhanced metabolic function.

### 2.5 3D printed liver discoids express mRNAs for ADME-associated genes

To characterize the transcriptional function of hepatocytes within printed tissue models, mRNA profiling studies were performed to assess the expression levels of ADME-associated genes following culture and maturation in the PEG microgel media. Relative expression levels were determined in printed discs of hepatocytes (H) and hepatocyte:HUVEC co-cultures (H:Hu) and compared to freshly-thawed hepatocyte cell suspensions (FTH), liver biopsy tissue, and HepatoPac – an established in vitro primary human hepatocyte co-culture model used in drug-toxicity studies (Hepregen Corporation). Relative expression levels of Phase 1 metabolizing enzymes and the Phase 2 enzyme, UGT1A1, are shown in Fig. 6a, and were found to be stable within printed discoids up to D21. Moreover, expression levels in discoids were mostly comparable to liver tissue and hepatocyte cell controls (Fig 6a). Differential expression analysis and hierarchical clustering was then performed among groups for a total of 115 ADME-associated gene probes and revealed that H:Hu and H prints clustered together, with the H:Hu D14 most similar to liver and hepatocyte cell (FTH) controls (Fig. 6b). Notably, both H:Hu and H prints were closer to liver controls than HepatoPac. These findings indicate that 3D printed discoids cultured in the PEG microgel media exhibit ADME function close to that of human liver and provide further evidence that HUVEC co-culture enhances hepatocellular functions.

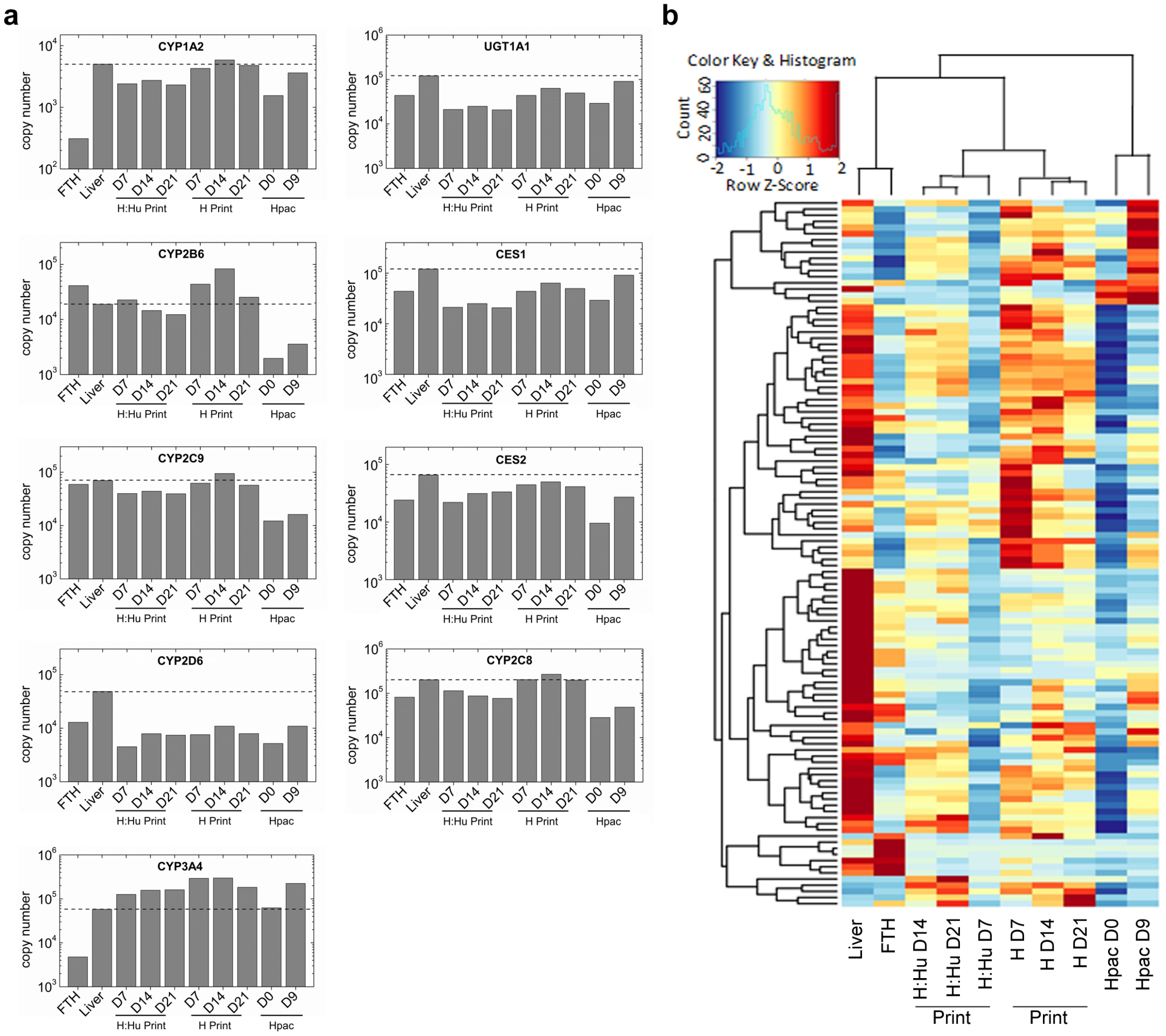

**Figure 6: ADME gene expression profile in 3D printed liver discoids.** (a) Relative gene expression of selected ADME-related genes in printed discoids . compared to freshly thawed hepatocytes (FTH), adult human liver donors (Liver) and human Hepatopac (Hpac). The dashed line indicates the expression level of FTH. (b) Heatmap visualizing relative expression levels of ADME and liver-function-associated genes (115 total) among groups. Data represent average values from 2 independent experiments. Note that 3D printed H:Hu co-cultures cluster closest to Liver and FTH controls.

**2.6 3D printed discoids retain CYP enzyme functions in microgel culture**

Since H:Hu co-cultures exhibited ADME gene expression profiles close to those of human liver and also exhibited higher levels of albumin and urea production than other candidate liver tissue models, we further studied small molecule metabolism in these co-cultures through the induction of liver phase I and II enzymatic activities after incubation with standard probe substrates. After 14 days in microgel culture, printed H:Hu discoids were incubated with a probe substrate cocktail to monitor the activity of the following enzymes: CYP1A2, CYP2C9, CYP2D6, CYP3A4, and UGT1A1. Samples were harvested following 0, 30, 60, 100, and 180 minutes of incubation, and metabolite concentration was measured using liquid chromatography-mass spectrometry (LC-MS; see methods). We performed the same measurements on freshly thawed suspension control samples, which serve as a benchmark for evaluating the *in vitro* induction of CYP enzymes[34]. Additionally, to determine if metabolic functions within printed H:Hu discoids could be enhanced, microgel cultures were supplemented with bioactive factors associated with the long-term maintenance of primary hepatocytes. The small molecule ROCK inhibitor, Y-27632, was selected based on its capacity to enhance the viability and long-term maintenance of hepatocytes [35], while Hepatocyte Growth Factor (HGF) is well established as a potent mitogen and trophic factor for primary hepatocytes[36]. Across both experimental groups and suspension control samples, we found metabolite concentration to increase over time, consistent with a linear scaling law when plotted on a log-log scale (Fig. 7a). This linear growth is further illustrated by re-scaling each metabolite concentration curve by its own slope, given by $\tilde{c}(t) = c(t)/m$, where the slope of each curve is estimated from $m = \langle c(t)/t \rangle_t$, and angle brackets correspond to an average over time points. This procedure results in a data collapse, as shown in Figure 7b.

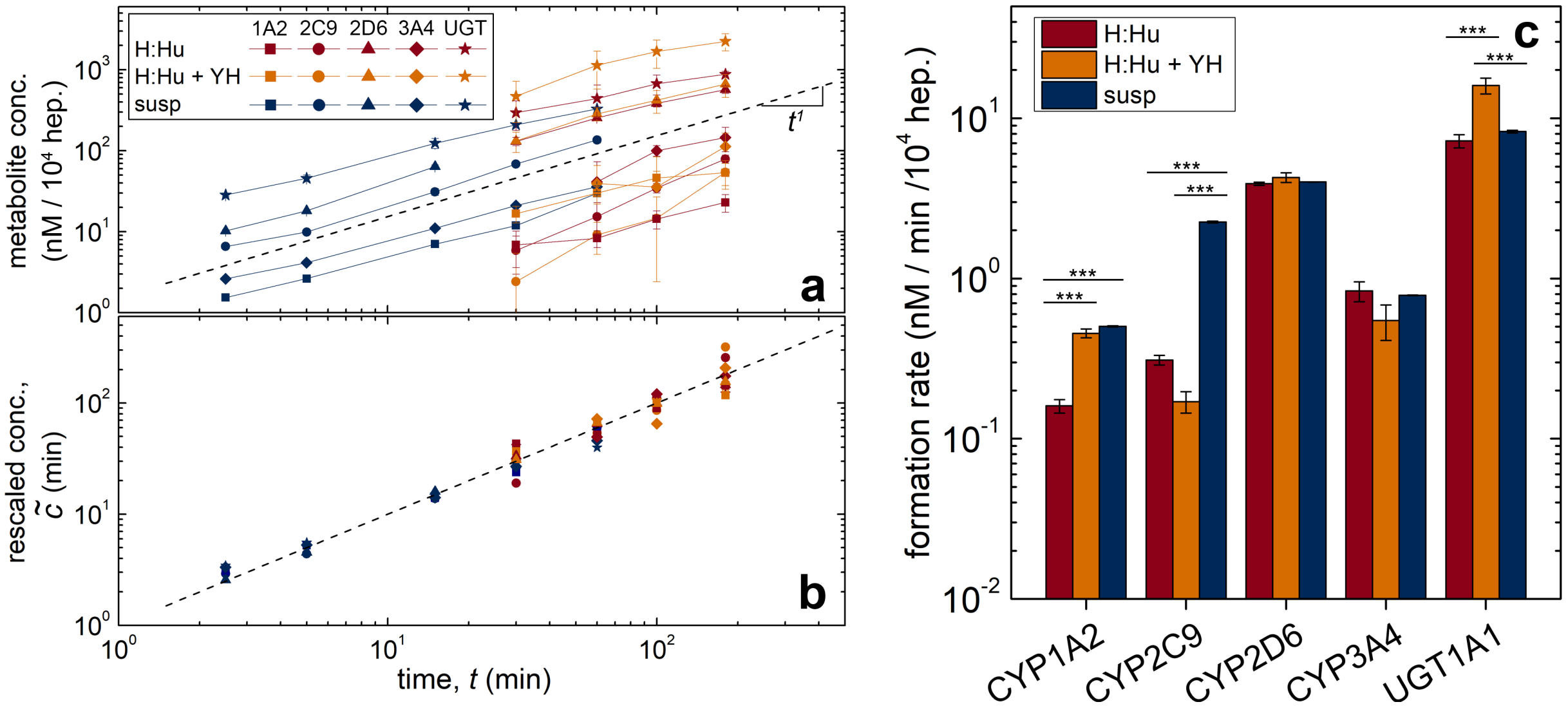

**Figure 7: Enzyme activity in 3D printed liver discoids.** (a) 3D printed liver tissue models were incubated with a probe substrate cocktail as described in section 4.7, and the activity of the CYP1A2, CYP2C9, CYP2D6, CYP3A4, and UGT1A1 enzymes was monitored by harvesting samples following 30, 60, 100, and 180 minutes of incubation and metabolite concentration measured using liquid chromatography-mass spectrometry (LC-MS). Suspension control samples (labeled as susp) were measured at earlier time-points. To determine if metabolic functions within printed H:Hu discoids could be enhanced, microgel cultures were supplemented with bioactive factors associated with the long-term maintenance of primary hepatocytes, such as the ROCK inhibitor, Y-27632, and Hepatocyte Growth Factor (labeled as H:Hu + YH). The metabolite concentrations across experimental groups and suspension control samples increased over time, consistent with a linear scaling law when plotted on a log-log scale. Individual datapoints represent mean ± standard deviation of measured concentrations per timepoint/group ($n$ = 5). (b) Each metabolite concentration profile was rescaled by its own slope and found to collapse on a master curve across multiple time points (dashed line represents $\tilde{c} = t$). (c) We performed linear regressions on each dataset to determine the metabolite formation rates based on the apparent linear increase of metabolite concentration with time (errorbars: standard errors of the best-fit slopes. *** $P < 0.001$.).

Based on the apparent linear increase of metabolite concentration with time, we performed linear regressions on each dataset to determine the metabolite formation rates. We found that for CYP2D6 and CYP3A4, the metabolite formation rate was not significantly different across the two experimental groups and the suspension control. For UGT1A1, the non-supplemented discoids and the suspension control were not significantly different, while supplementation increased the metabolite formation rate by a factor of about 2.5. Metabolite formation by the non-supplemented discoids was significantly lower than suspension control for CYP1A2 and CYP2C9. Supplementation boosted CYP1A2 activity but did not significantly change CYP2C9 activity (Fig. 7c). We note that the metabolite formation rates of freshly thawed primary suspension controls do not represent target performance metrics of 3D printed liver tissue models but instead serve as sensitivity checks on the LC-MS measurements. Still, it is noteworthy that after two weeks in culture after printing, the majority of formation rates with our discoid model are the same or higher than those with freshly thawed suspension cells. Motivated by these results that indicate the potential benefits of supplementation on liver enzyme activity, we tested whether Y-27632/HGF supplementation influenced albumin or urea synthesis. Indeed, we found that albumin and urea synthesis rates significantly increased 2-fold and 1.5-fold, respectively, compared to non-supplemented cultures (Supplementary Fig. S2), indicating that these factors may be used to further enhance liver functions in 3D printed discoids cultured in microgel medium.

## 3. Discussion and Conclusions

Here we have described an approach for reproducibly manufacturing human liver tissue models that exhibit key functions like sustained albumin and urea synthesis and ADME gene transcriptional profiles that fall within the range of human liver. Functionality of the 3D printed liver tissue models was also demonstrated by the induction of liver phase I and II enzymatic activities after incubation with standard probe substrates. The impacts of size and geometry on liver tissue model function were investigated by varying print trajectory paths to generate H:Hu constructs having different diameters and shapes. Overall, the generation of smaller tissue models led to higher levels of hepatocyte function within our culture system. Increasing the construct diameter of discoids reduced urea synthesis rates, while both albumin and urea production were markedly lower in printed spheroids compared to discs. Corresponding differences in cell organization were noted between print geometries; discoids were more compact with greater cell-cell adhesion, while spheroids were less rigid with loosely arranged cells. Consistent with these observations, cell-cell communication through gap junction complexes is known to be critical to maintaining liver homeostasis, with gap junctions playing a prominent role in maintaining a differentiated phenotype, including regulation of liver-specific processes such as albumin secretion[37, 38]. We anticipate that hepatocyte functions within larger tissue models, including spheroids, may be improved by adjusting print parameters to establish conditions for greater cell-cell cohesion following extrusion. This may be achieved through alterations of microgel composition, print nozzle translation speed, extrusion rate, and cell or collagen concentrations.

Small-scale tissue models like the liver discoids developed here often need to be cultured over long periods, during which liquid media must be exchanged for tissue feeding, dosing, or assaying[20, 23]. Fluid exchange was performed by hand in our work, and beyond the potential labor-saving benefits of automated perfusion, it is possible that tissue function could be further enhanced by employing perfusion. For example, liver tissue models that are a few cell layers thick cultured with microfluidic perfusion chambers have been shown to exhibit higher viability, oxygen saturation, and synthesis rates of albumin and urea than their static counterparts[22]. A crucial advantage of the embedded 3D bioprinting approach we took here is that the printing support medium serves a second purpose as the culture environment; structures can be fabricated directly into a 3D culture media that supports them mechanically and metabolically. Unfortunately, perfusion through previously established embedded 3D printing support media is prohibitive; these materials are typically large-scale polymer networks or packs of irregularly shaped soft micro-particles with nano-scale mesh sizes that pack conformally and exhibit little interstitial pore space[29]. By contrast, the PEG microgels used here are spherical, so micron-scale pores form when they are loosely packed together[39]. We tested whether this extra pore-space enhances permeability, finding that the PEG microgels used here are 20 times more permeable than previously used embedding materials or continuous hydrogels (Supplementary Fig. S1). Thus, while we found perfusion was unnecessary to achieve excellent tissue function, our approach is compatible with perfusion, and we envision it being employed in applications where perfusion is needed.

Our choice to make discoids embedded in porous microgel packs enabled us to side-step the need to perfuse structures made from large, dense cell populations for their maintenance and maturation, which is one of the most frequently cited challenges in developing engineered 3D cellular constructs[40]. Perfusable architectures have been made by 3D printing biocompatible scaffolds with large gaps that facilitate cell infiltration, and the driven flow of culture media and cell waste[41, 42]. Perfusable fluid channels have been built into scaffolds to support dense cell populations[43]. Inventive approaches have been developed using embedded 3D printing instead of scaffolding. For example, sacrificial materials have been printed into networks embedded in cell-laden biomaterials where a sacrificial network is evacuated and coated with endothelial cells, creating vascularized cell populations with controlled spatial distributions[24, 25]. We believe our findings represent a path toward developing functional tissue models that do not require vascularization or perfusion.

While our results demonstrate that packed PEG microgels serve as an excellent support medium for embedded 3D bioprinting and are suitable for long-term culture and assaying of tissue models, it is likely that multiple cell types found within native tissue must be incorporated to replicate the broad spectrum of liver specific functions. In our model, co-culture of hepatocytes and HUVECs enhanced hepatocyte function, exhibiting increased albumin and urea synthesis rates and an ADME gene expression profile that more closely resembled native liver tissue than an established 2D model. HUVECs have been previously reported to enhance hepatocyte functions in a variety of co-culture systems[44-46], where they may mimic the hepatocyte-stabilizing functions of liver sinusoidal endothelial cells[47]. The secretion of soluble factors and the formation of heterotypic cell-cell contacts have been implicated in other co-culture systems using bovine aortic endothelial cells[48, 49], which may also underly the enhanced performance we find with the addition of HUVECs. In contrast to HUVECs, cholangiocyte incorporation did not enhance hepatocyte functions in our tissue model, causing a significant reduction in both albumin and urea production rates. Diminished hepatocyte functions have been reported in cholangiocyte co-culture with iPSC derived hepatocytes[50] and may be attributed to cholangiocyte overgrowth and cell activation. These observations warrant further investigation of the cholangiocyte phenotype in 3D printed tissue models and indicate the need to optimize cell ratios and culture conditions for more accurate modeling of native liver function. Our findings on enzymatic activity also provide a basis for further investigation of growth factor and small molecule combinations in preserving the hepatocellular functions of 3D printed microtissues. Additional approaches for retaining CYP functions in long-term hepatocyte culture include the incorporation of cell/matrix components[51, 52], the application of fluid flow[53, 54], and media supplementation with bioactive factors[35, 55], which may be useful in future work to further enhance the human liver tissue model developed here.

## Materials and Methods

### 4.1 PEG microgel synthesis

Polyethylene glycol (PEG) microgels were synthesized by inverse emulsion polymerization where an aqueous solution of 99 mol% poly(ethylene glycol) methyl ether acrylate (Mn = 480g/mol), 1 mol% poly(ethylene glycol) diacrylate (Mn = 700g/mol) and 0.15% (w/w) of ammonium persulfate was prepared. The aqueous solution was dripped into an organic solution of 0.7% (w/w) polyglycerol polyricinoleate (PGPR, a commercial surfactant) in dodecane and homogenized at 8000 rpm for 5 minutes. The mixture was placed in a round bottom flask in an ice bath, and dissolved oxygen was purged by a nitrogen stream for 1 hour. After removing the mixture from the ice bath, 1,2-Di (dimethyl amino) ethane was added at a final concentration of 0.5 % (w/w) while stirring in a nitrogen atmosphere for 1 hour. The reaction was completed by exposing the solution to air while stirring for 30 min. The resulting microgel solution was washed with solvents by vigorous shaking and centrifugation at 4000 x gravity (g) for 15 minutes, followed by the removal of supernatant; these steps were performed multiple times in ethanol, followed by ultrapure water, and then phosphate-buffered saline (PBS) to remove unreacted components and residues. The microgels swollen in PBS were sterilized by autoclaving at 121℃ and 15 psi. Before printing, the gel was washed with basal Williams E medium without phenol red. This was followed by a fluid exchange with hepatocyte plating media (ThermoFisher, #CM9000). Finally, 25% (v/v) of plating media was added, and the microgels were swollen in the media overnight.

**4.2 PEG microgel characterization**

Material properties of the PEG microgel packs are measured by conducting shear rheology experiments on an Anton Paar MCR 702 rheometer with 25 mm and 50 mm parallel plate geometries with a measuring gap of 1mm. Elastic and viscous shear moduli of the microgel packs are measured in the linear deformation regime using small amplitude frequency sweeps at 2% strain amplitude over a frequency range of 0.1-10Hz. The yielding of the microgel packs under persistent shear is tested using unidirectional shear tests, ramping the shear rate from 0.001 to 100 $s^{-1}$ while measuring the shear stress.

**4.3 Cells culture and 3D bioprinting**

Liver tissue model biofabrication was performed by printing various cell structures in microgel media swollen with hepatocyte growth medium using a custom-fabricated precision 3D printer (DEKA Research and Development Corp.). For biofabrication quality studies, hepatocytes (HepaRG, ThermoFisher #HPRGC10) fluorescently labeled with CellTracker fluorescent probes (CMFDA, ThermoFisher #C7025) were used. For tissue function studies, primary human hepatocytes (ThermoFisher, #HMCPTS) were directly thawed from cryopreserved stocks, according to the vendor's instructions, counted and resuspended into a minimum volume (~200 µL) of hepatocyte plating medium (ThermoFisher, #CM9000) supplemented with 1 mg/mL type I collagen solution (Advanced Biomatrix, Nutragen #5010) at a concentration of ~5 x $10^7$ cells/mL. The cell-collagen solution was then transferred into a 250µL Hamilton gastight syringe with a sterile, blunt-tip 30-gauge luer-lock needle for printing into microgel medium in standard 12-well or 96-well tissue culture plates, and 3D constructs were generated using GCODE trajectories programmed in MATLAB.

For co-culture experiments, human umbilical vein endothelial cells (HUVEC; Lonza #C2519AS) and primary human cholangiocytes (CelProgen, #36755-11), were first maintained in monolayer cultures in endothelial (Lonza, #CC3156) or primary cholangiocyte (#M36755-12S) culture medium respectively, and then harvested by gentle trypsinization prior to printing. The cells were premixed with primary hepatocytes at defined ratios in collagen solution prior to syringe loading and printing. After printing, microgel medium was overlaid with an equal volume of hepatocyte plating medium, and after 3 days the supernatant medium was replaced with hepatocyte maintenance medium (ThermoFisher, #CM4000) for extended culture. Discoid areas shown in Fig. 4d were determined from images of tissues harvested after 10 days in culture. Using Image J software (NIH, Bethesda), mean areas ($mm^2$) were calculated and plotted for each culture group.

### 4.4 Measurement of albumin and urea synthesis

For each measurement, supernatant media from microgel cultures was exchanged with fresh hepatocyte maintenance medium 24 h prior to collection. All samples were stored at -80°C prior to assay. Albumin levels were measured by sandwich ELISA using the Human Albumin ELISA kit (Bethyl Laboratories, #E88-129), according to the manufacturer's protocol. Urea levels were determined from the same samples using the Stanbio Urea Nitrogen (BUN) kit (Stanbio Labs). Synthesis rates were calculated per $10^6$ cells following cell number estimation.

### 4.5 Determination of cell viability and number

Cell numbers were estimated from DNA content, determined by PicoGreen assay of harvested microtissues. Samples were digested with papain (300 μg/mL), freeze-thawed, and incubated with Quant-iT™ PicoGreen® dsDNA reagent (Molecular Probes, #P7581) according to the manufacturer's protocol. Following a 5-minute incubation, samples were measured on a fluorescence microplate reader, and DNA concentrations calculated from a standard curve.

### 4.6 Histology and Immunohistochemistry

Microtissues were harvested from the microgels, washed in PBS, fixed in 4% PFA for 30 min at room temperature, and embedded in 0.7% agarose for ease of handling. After dehydration in graded alcohols, the samples were paraffin embedded and sectioned to 5μm. Following deparaffinization and rehydration, sections were stained with H&E. For immunohistochemistry sections were photobleached overnight in 5% $H_20_2$ solution to reduce autofluorescence, blocked with serum and incubated overnight with antibodies to human albumin (ThermoFisher, #A80-229A), and CD31 (Abcam, #ab28364) for detection of hepatocytes and HUVECs, respectively. Following incubation with anti-goat APC-conjugated and anti-rabbit PE-conjugated antibodies (ThermoFisher) sections were nuclear stained with Hoescht 33342 (Invitrogen) and analyzed by fluorescence microscopy using an EVOS M5000 Imaging System (ThermoFisher).

**4.7 RNA isolation and gene array profiling**

Total RNA was extracted from pooled printed microtissues (3-4) or freshly thawed primary hepatocytes (FTH) from a matched lot, following lysis in TRIzol reagent (Invitrogen), phenol-chloroform extraction, and isopropanol precipitation. Column-based purification was then performed using the RNeasy MicroKit (Qiagen). To provide additional reference controls for ADME gene expression, total RNA was also obtained from human liver biopsy tissue, and human Hepatopac cultures, providing additional controls. Equal amounts of input total RNA were used for gene profiling using TaqMan Array panels (Applied Biosystems) containing >100 probes for ADME-associated genes, according to the Manufacturer's protocol (see supplementary Table S1 for full gene list). Raw CT Values were converted to Theoretical Copy #: $2^{(40\text{- CT value of target gene})}$ /10K.

For differential gene expression analysis, a batch effect correction using the empirical Bayes strategy was first performed by using the sample groups as covariates [56]. Data was also normalized to minimize differences among replicates [57]. For hierarchical clustering, the complete linkage method was used, with Pearson's correlation analysis as the distance metric. Housekeeping genes were removed from the dataset, and the normalized data was used to plot the heatmaps. A total of 115 genes are present in the analysis.

**4.8 Drug metabolism assay**

*Drug incubation* - Printed microtissues were assayed for enzyme function following 14 d of microgel culture. Samples were harvested from the microgels and transferred to standard 96-well plates for drug incubation at 37 °C with a premade probe substrate cocktail comprising Phenacetin (100 mM, Sigma-Aldrich), Diclofenac (25 mM, Sigma-Aldrich), Dextromethorphan (50 mM, Alfa Aesar), Nifedipine (15 mM, Sigam-Aldrich), and 7-hydroxycoumarin (10 mM, Sigma-Aldrich). The cocktail was used to monitor metabolite formation in response to the activity of the following enzymes: CYP1A2, CYP2C9, CYP2D6, CYP3A4, and UGT1A1. Substrate concentrations were based on historical data and have demonstrated linear formation kinetics across the specified time course. Samples were harvested following 0, 30, 60, 100, and 180 minutes of incubation, mixed with acetonitrile containing an internal standard cocktail of labetalol and imipramine (Cerilliant), and stored at -80 °C until analysis. A matched lot of freshly thawed hepatocyte suspension controls were treated in parallel to compare enzyme functions.

*LC-MS analysis* – Samples were thawed on wet ice. Samples were centrifuged in an Eppendorf table-top centrifuge (model 5425) at 3220 *xg* at 4 °C for 15 minutes to pellet precipitate. Following centrifugation, 100 µL of supernatant was transferred to a 96-well high recovery analysis plate. 100 µL of water was added to supernatant and briefly vortexed. LC-MS/MS analysis was performed on a Thermo Scientific (Waltham, Ma) LX-2 UPLC system with a Leap Autosampler interfaced to an Applied Biosystems/MSD Sciex (Framingham, MA) API-6500 mass spectrometer utilizing a turbo ion spray interface in both positive and negative modes. Separation was achieved using an Acquity UPLC HSS T3 column (1.8 µm, 2.1 x 50 mm) with a mobile phase consisting of 0.1% formic acid in water (solvent A) and 0.1% formic acid in acetonitrile (solvent B) at a flow rate of 0.75 mL/min. The LC gradient began at 5% B and held for 0.25 minutes then changed to 95% B over 1.5 min, held at 95% B for 0.42 minutes, and then returned to 5% B for 0.83 minutes. Analyte response was measured by multiple reaction monitoring (MRM) of transitions unique to each substrate's metabolite: *m/z* 312 to *m/z* 231 for hydroxydiclofenac, *m/z* 258 to *m/z* 157 for dextrorphan, *m/z* 152 to *m/z* 110 for acetominophen, *m/z* 345.1 to *m/z* 284.3 for oxidized nifedipine, and *m/z* 337 to *m/z* 175 for 7-hydroxycoumarin glucuronide with *m/z* 329 to *m/z* 162.1 for labetalol internal standard and *m/z* 281.3 to *m/z* 193.1 for imipramine internal standard.

**Acknowledgements**

The authors thank Anton Paar for the use of the Anton Paar 702 rheometer through their VIP academic research program. Research was sponsored by the Office of the Secretary of Defense and was accomplished under Agreement Number W911NF-17-3-0003. The views and conclusions contained in this document are those of the authors and should not be interpreted as representing the official policies, either expressed or implied, of the Office of the Secretary of Defense or the U.S. Government. The U.S. Government is authorized to reproduce and distribute reprints for Government purposes notwithstanding any copyright notation herein.

**Supporting Information**

Permeability of packed microgel media; Functional activity of 3D printed liver microtissues following media supplementation with Y-2732 and HGF; Gene list of TaqMan Array panel; Figure S1; Figure S2; Table S1.

# Supplementary Information

# A Functional Human Liver Tissue Model: 3D Bioprinted Co-culture Discoids

**Vignesh Subramaniam[1,†], Carolina Abrahan[2,†], Brett R. Higgins[3], Steven J. Chisolm[1], Baleigh Sweeney[1], Senthilkumar Duraivel[4], Leandro Balzano-Nogueira[5], Glyn D. Palmer[2,*], Thomas E. Angelini[1,6,7,*]**

1 Department of Mechanical and Aerospace Engineering, Herbert Wertheim College of Engineering, University of Florida, Gainesville, Florida, United States of America

2 Department of Orthopaedic Surgery and Sports Medicine, College of Medicine, University of Florida, Gainesville, Florida, United States of America

3 Department of Molecular Medicine, Morsani College of Medicine, University of South Florida, Tampa, Florida 33612, United States of America

4 Department of Materials Science and Engineering, Cornell University, Ithaca, New York, United States of America

5 Department of Pathology, Immunology and Laboratory Medicine, Diabetes Institute, College of Medicine, University of Florida, Gainesville, Florida, United States of America.

6 Department of Materials Science and Engineering, Herbert Wertheim College of Engineering, University of Florida, Gainesville, Florida, United States of America

7 J. Crayton Pruitt Family Department of Biomedical Engineering, Herbert Wertheim College of Engineering, University of Florida, Gainesville, Florida, United States of America

† The authors contributed equally to this work.

* Corresponding authors. Email: t.e.angelini@ufl.edu, palmegd@ortho.ufl.edu

**S1. Permeability of packed microgel media**

As described in Materials and Methods, we employ an emulsion-based polymerization method to create a perfusable 3D bioprinting medium made from spherical microgels. Previous work on microgel-based media for 3D printing and culture employed precipitation polymerization to synthesize the gels, producing irregularly shaped particles that pack together densely, resulting in small pores between the microgels [1, 2]. By contrast, spherical particles of approximately the same size as their irregularly shaped counterparts are expected to exhibit larger pores between the microgels, resulting in a higher permeability medium. We conducted a series of fluid-flow tests through vertical columns to test whether support media made from packed spherical microgels are more permeable to fluid flow than media made from irregularly shaped gels. Mesh screens were fixed to the lower ends of the columns to hold microgel packs in place during the tests. Microgel dispersions were pipetted into the columns and allowed to settle under gravity for several hours. The base of each column was immersed in a wide, shallow basin capable of collecting large fluid volumes with negligible changes in fluid height. Each column was filled with water with gentle pipetting without disturbing the microgel pack. We performed time-lapse imaging to quantify the instantaneous fluid flow rate and pressure throughout the experiments as the water flowed through the microgels. Ignoring the fluid level rise in the basin, the height of the fluid level in the column is expected to drop exponentially, given by $h(t) = Ce^{-t/\tau}$, with a characteristic decay time of $\tau = L/(k\rho g)$, where $L$ is the height of the microgel pack, $k$ is the microgel pack permeability, $r$ is the mass density of water, and $g$ is the acceleration due to gravity. The resistance to fluid flow provided by the microgel pack is controlled by $L$ and $k$; the driving force behind fluid flow is controlled by $r$ and $g$. (Fig S1a). Plotting our h(t)/C measurements on a semilog-Y plot, we find the data lay on straight lines, consistent with exponential decay. Thus, we fit an exponential decay model to each dataset to determine the sample permeability, $k$.

To investigate the role of particle size in packed microgel permeability, we performed these flow tests on packs of spherical microgels having a wide range of diameters, $d$, between approximately 10 and 150 μm. These include three different PEG microgel formulations and two different commercially available microgels made from polyacrylamide (Biorad). Plotting $k$ versus $d$ on a log-log scale, we see that $k$ scales approximately like $d^2$ (Fig. S1b). This scaling is consistent with the Kozeny-Carman relation, given by

$$k = \frac{f^2 \varepsilon^3 d^2}{150(1-\varepsilon)^2 \eta},$$

where $f$ is the sphericity of the particles, $\varepsilon$ is the porosity of the particle pack, and $h$ is the fluid viscosity [3]. In our case, $f = 1$ since the particles are spherical. This relationship lies close to the data, assuming the viscosity of water and $\varepsilon$ of 0.32, close to the value for loosely packed spheres [4, 5]. To test how particle shape influences the permeability of microgel packs, we plot the data from samples of irregularly shaped methacrylic acid particles. Here, for particles having an average diameter of approximately 10 µm, we find that the permeability of the irregularly shaped microgel pack is between 15 and 40 times smaller than its spherical counterpart. Thus, 3D printing and culturing cellular structures in support media made from spherical microgels is expected to dramatically enhance the permeability of the culture environment, correspondingly reducing the pressure differential needed for the perfusion of liquid media.

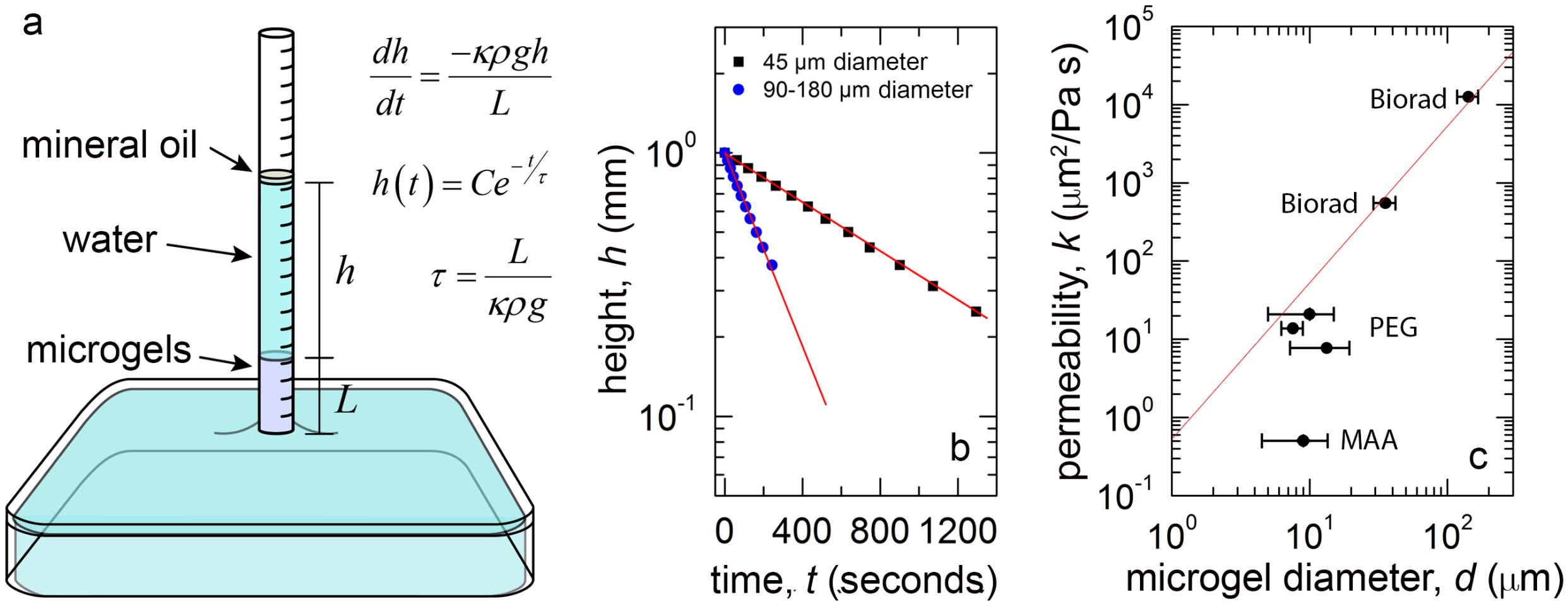

**Figure S1: Microgel permeability.** We conducted a series of fluid flow tests through vertical columns to determine whether support media made from spherical microgels would be suitable for 3D printing applications and perfused culture conditions. (a.) The permeability of loosely packed microgel samples were measured using a gravity-driven flow apparatus in which the height of a fluid column was measured over time. (b.) We found that the fluid column height, h, falls exponentially with time given by $h(t) = Ce^{-t/\tau}$, with a characteristic decay time of $\tau = L/(k\rho g)$, where $L$ is the height of the microgel pack, $k$ is the microgel pack permeability, $r$ is the mass density of water, and $g$ is the acceleration due to gravity. To investigate the role of particle size in packed microgel permeability, we performed these flow tests on packs of spherical microgels having a wide range of diameters, $d$, between approximately 10 and 150 µm. These include three different PEG microgel formulations and two different commercially available microgels made from polyacrylamide (Biorad). Plotting $k$ versus $d$ on a log-log scale, we see that $k$ scales approximately like $d^2$, consistent with the Kozeny-Carman relation. (c.) To test how particle shape influences the permeability of microgel packs, we plot the data from samples of irregularly shaped methacrylic acid particles and find that for particles having an average diameter of approximately 10 µm, the permeability of the irregularly shaped microgel pack is between 15 and 40 times smaller than its spherical counterpart.

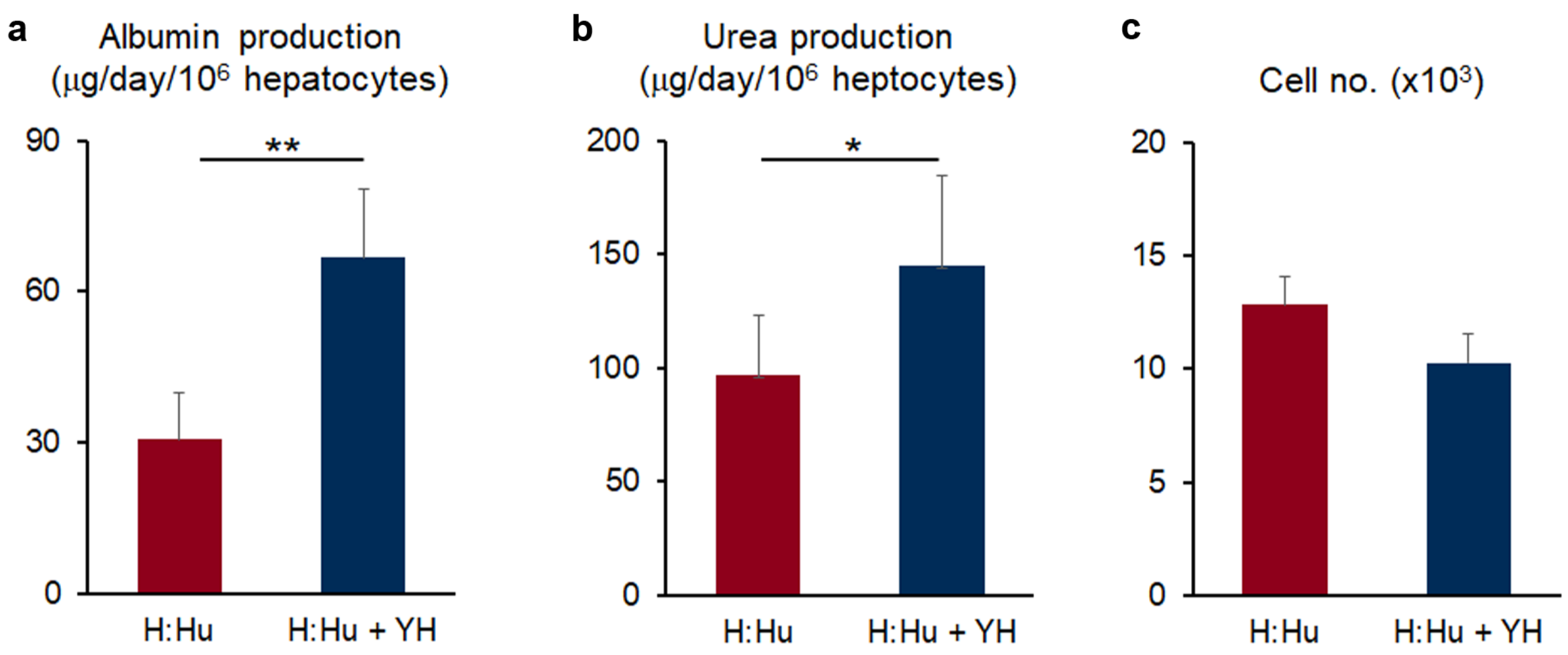

**Figure S2: Functional activity of 3D printed liver microtissues following media supplementation with Y-2732 and HGF.** Heptatocyte-HUVEC (H:Hu) discoids were cultured for 14 d with or without supplementation of the LLS media with HGF (20 ng/mL) and the ROCK inhibitor, Y-2732 (10 mM). (a) Mean cell number per printed construct. (b) Albumin and (c) urea synthesis rates, adjusted to hepatocyte cell number for each group. Values are mean + SD for each group (n=5). * $P < 0.05$; ** $P < 0.05$. The results show that YH supplementation significantly enhances albumin and urea synthesis rates in 3D printed H:Hu co-cultures without affecting cell growth/viability. These findings are consistent with YH-mediated increases in CYP functions, shown in Figure 7, and suggest that their supplementation can broadly enhance hepatocellular functions of 3D printed microtissues cultured in LLS media.

## Table S1. Gene list of TaqMan Array panel for ADME expression profiling of 3D printed liver tissue models.

| Gene Symbol | Gene Name |
|---|---|
| 18S | RNA18S1 |
| ABCB1 | ATP Binding Cassette Subfamily B Member 1 |
| ABCB11 | ATP Binding Cassette Subfamily B Member 11 |
| ABCB6 | ATP Binding Cassette Subfamily B Member 6 |
| ABCC2 | ATP Binding Cassette Subfamily C Member 2 |
| ABCC3 | ATP Binding Cassette Subfamily C Member 3 |
| ABCC4 | ATP Binding Cassette Subfamily C Member 4 |
| ABHD4 | Abhydrolase Domain Containing 4, N-Acyl Phospholipase B |
| ACOX1 | Acyl-CoA Oxidase 1 |
| AKR1C3 | Aldo-Keto Reductase Family 1 Member C3 |
| ALAS1 | 5'-Aminolevulinate Synthase 1 |
| ATPIF1 | ATP Synthase Inhibitory Factor Subunit 1 |
| BCAT2 | Branched Chain Amino Acid Transaminase 2 |
| BLVRB | Biliverdin Reductase B |
| CBR1 | Carbonyl Reductase 1 |
| CES1 | Carboxylesterase 1 |
| CES2 | Carboxylesterase 2 |
| CLCF1 | Cardiotrophin Like Cytokine Factor 1 |
| COPS4 | COP9 Signalosome Subunit 4 |
| COPS5 | COP9 Signalosome Subunit 5 |
| CYP1A1 | Cytochrome P450 Family 1 Subfamily A Member 1 |
| CYP1A2 | Cytochrome P450 Family 1 Subfamily A Member 2 |
| CYP1B1 | Cytochrome P450 Family 1 Subfamily B Member 1 |
| CYP2B6 | Cytochrome P450 Family 2 Subfamily B Member 6 |
| CYP2C19 | Cytochrome P450 Family 2 Subfamily C Member 19 |
| CYP2C8 | Cytochrome P450 Family 2 Subfamily C Member 8 |
| CYP2C9 | Cytochrome P450 Family 2 Subfamily C Member 9 |
| CYP2D6 | Cytochrome P450 Family 2 Subfamily D Member 6 |
| CYP2E1 | Cytochrome P450 Family 2 Subfamily E Member 1 |
| CYP3A4 | Cytochrome P450 Family 3 Subfamily A Member 4 |
| CYP3A43 | Cytochrome P450 Family 3 Subfamily A Member 43 |
| CYP3A5 | Cytochrome P450 Family 3 Subfamily A Member 5 |
| CYP4A11 | Cytochrome P450 Family 4 Subfamily A Member 11 |
| CYP4A22 | Cytochrome P450 Family 4 Subfamily A Member 22 |
| CYP7A1 | Cytochrome P450 Family 7 Subfamily A Member 1 |
| CYP8B1 | Cytochrome P450 Family 8 Subfamily B Member 1 |
| DDX47 | DEAD-Box Helicase 47 |
| ENTPD5 | Ectonucleoside Triphosphate Diphosphohydrolase 5 |
| EPHX1 | Epoxide Hydrolase 1 |
| EPHX2 | Epoxide Hydrolase 2 |
| FABP1 | Fatty Acid Binding Protein 1 |
| FTH1 | Ferritin Heavy Chain 1 |
| GAPDH | Glyceraldehyde-3-Phosphate Dehydrogenase |
| GPLD1 | Glycosylphosphatidylinositol Specific Phospholipase D1 |
| GPX1 | Glutathione Peroxidase 1 |
| GPX2 | Glutathione Peroxidase 2 |
| GPX3 | Glutathione Peroxidase 3 |
| GSTA1 | Glutathione S-Transferase Alpha 1 |
| GSTA2 | Glutathione S-Transferase Alpha 2 |
| GSTK1 | Glutathione S-Transferase Kappa 1 |
| GSTM3 | Glutathione S-Transferase Mu 3 |
| GSTP1 | Glutathione S-Transferase Pi 1 |
| GSTZ1 | Glutathione S-Transferase Zeta 1 |
| GUSB | Glucuronidase Beta |
| HADHA | Hydroxyacyl-CoA Dehydrogenase Trifunctional Multienzyme Complex Subunit Alpha |
| HMOX1 | Heme Oxygenase 1 |
| HNRNPUL1 | Heterogeneous Nuclear Ribonucleoprotein U Like 1 |
| INPP5A | Inositol Polyphosphate-5-Phosphatase A |
| IRAK1 | Interleukin 1 Receptor Associated Kinase 1 |
| KEAP1 | Kelch Like ECH Associated Protein 1 |
| ME1 | Malic Enzyme 1 |
| MGST2 | Microsomal Glutathione S-Transferase 2 |
| MGST3 | Microsomal Glutathione S-Transferase 3 |
| MKI67 | Marker Of Proliferation Ki-67 |
| NMT1 | N-Myristoyltransferase 1 |
| NQO1 | NAD(P)H Quinone Dehydrogenase 1 |
| PIR | Pirin |
| POR | Cytochrome P450 Oxidoreductase |
| PPAP2B | Phospholipid Phosphatase 3 |
| PPARA | Peroxisome Proliferator Activated Receptor Alpha |
| PPARG | Peroxisome Proliferator Activated Receptor Gamma |
| PSMA6 | Proteasome 20S Subunit Alpha 6 |
| PSMB1 | Proteasome 20S Subunit Beta 1 |
| PSMB2 | Proteasome 20S Subunit Beta 2 |
| PSMB4 | Proteasome 20S Subunit Beta 4 |
| PSMB7 | Proteasome 20S Subunit Beta 7 |
| PSMC4 | Proteasome 26S Subunit, ATPase 4 |
| PSMD1 | Proteasome 26S Subunit, Non-ATPase 1 |
| PSMD11 | Proteasome 26S Subunit, Non-ATPase 11 |
| PSMD14 | Proteasome 26S Subunit, Non-ATPase 14 |
| PSMD2 | Proteasome 26S Subunit Ubiquitin Receptor, Non-ATPase 2 |
| PSMD3 | Proteasome 26S Subunit Ubiquitin Receptor, Non-ATPase 3 |
| PSMD7 | Proteasome 26S Subunit Ubiquitin Receptor, Non-ATPase 7 |
| PUM1 | Pumilio RNA Binding Family Member 1 |
| RAB35 | RAB35, Member RAS Oncogene Family |
| RAN | RAN, Member RAS Oncogene Family |
| RCHY1 | Ring Finger And CHY Zinc Finger Domain Containing 1 |
| REXO2 | RNA Exonuclease 2 |
| RGN | Regucalcin |
| RNASE4 | Ribonuclease A Family Member 4 |
| SERPIND1 | Serpin Family D Member 1 |
| SETD4 | SET Domain Containing 4 |
| SLC16A10 | Solute Carrier Family 16 Member 10 |
| SLC22A6 | Solute Carrier Family 22 Member 6 |
| SLC6A2 | Solute Carrier Family 6 Member 2 |
| SLCO1B1 | Solute Carrier Organic Anion Transporter Family Member 1B1 |
| SLCO1B3 | Solute Carrier Organic Anion Transporter Family Member 1B3 |
| SOD1 | Superoxide Dismutase 1 |
| SOD2 | Superoxide Dismutase 2 |
| SOD3 | Superoxide Dismutase 3 |
| SRRM1 | Serine And Arginine Repetitive Matrix 1 |
| SRXN1 | Sulfiredoxin 1 |
| THRSP | Thyroid Hormone Responsive |
| TKT | Transketolase |
| TLK2 | Tousled Like Kinase 2 |
| TMED4 | Transmembrane P24 Trafficking Protein 4 |
| TMEM183A | Transmembrane Protein 183A |
| TRIM37 | Tripartite Motif Containing 37 |
| TXNRD1 | Thioredoxin Reductase 1 |
| UFD1L | Ubiquitin Recognition Factor In ER Associated Degradation 1 |
| UGT1A1 | UDP Glucuronosyltransferase Family 1 Member A1 |
| UGT1A3 | UDP Glucuronosyltransferase Family 1 Member A3 |
| UGT1A4 | UDP Glucuronosyltransferase Family 1 Member A4 |
| UGT1A5 | UDP Glucuronosyltransferase Family 1 Member A5 |
| UGT1A6 | UDP Glucuronosyltransferase Family 1 Member A6 |
| UGT1A7 | UDP Glucuronosyltransferase Family 1 Member A7 |
| UGT1A8 | UDP Glucuronosyltransferase Family 1 Member A8 |
| UGT1A9 | UDP Glucuronosyltransferase Family 1 Member A9 |
| UGT2B10 | UDP Glucuronosyltransferase Family 2 Member B10 |
| UGT2B11 | UDP Glucuronosyltransferase Family 2 Member B11 |
| UGT2B15 | UDP Glucuronosyltransferase Family 2 Member B15 |
| UGT2B17 | UDP Glucuronosyltransferase Family 2 Member B17 |
| UGT2B4 | UDP Glucuronosyltransferase Family 2 Member B4 |
| UGT2B7 | UDP Glucuronosyltransferase Family 2 Member B7 |
| USP5 | Ubiquitin Specific Peptidase 5 |
| VCP | Valosin Containing Protein |
| ZFAND2A | Zinc Finger AN1-Type Containing 2A |
| ZWINT | ZW10 Interacting Kinetochore Protein |